\documentclass[a4j,11pt]{article}
\usepackage{amsthm}
\usepackage{amsmath}
\usepackage{amssymb}
\usepackage{amsfonts}
\usepackage{wrapfig}
\usepackage[dvipdfm]{graphicx,color}
\usepackage{float}
\usepackage{mathrsfs}
\usepackage{comment}
\usepackage{latexsym}

\theoremstyle{plain}
\newtheorem{col}{Corollary}
\newtheorem{Remark}{Remark}
\newtheorem{thm}{Theorem}

\newcommand{\argmax}{\mathop{\rm arg~max}\limits}

\renewcommand{\thesection}{\arabic{section}}

\begin{document}

\title{Second-order unbiased naive estimator of mean squared error for EBLUP in small-area estimation}
\author{Masayo Yoshimori Hirose\\ The Institute of Statistical Mathematics}
\date{\hfill}
\maketitle

\begin{abstract}
\indent An empirical best linear unbiased prediction (EBLUP) estimator is utilized for efficient inference in small-area estimation.
To measure its uncertainty, we need to estimate its mean squared error (MSE) since the true MSE cannot generally be derived in a closed form.
The {\it naive MSE estimator}, one of the estimators available for small-area inference, is unlikely to be chosen, since it does not achieve the desired asymptotic property, namely second-order unbiasedness, although it maintains strict positivity and tractability. Therefore, users tend to choose the second-order unbiased MSE estimator.
In this paper, we seek a new adjusted maximum-likelihood method to obtain a naive MSE estimator that achieves the required asymptotic property. To obtain the result, we also reveal the relationship between the general adjusted maximum-likelihood method for the model variance parameter and the general functional form of the second-order unbiased, and strictly positive, MSE estimator.
We also compare the performance of the new method with that of the existing naive estimator through a Monte Carlo simulation study. The results show that the new method remedies the underestimation associated with the existing naive estimator.
\end{abstract}

{\bf Keywords}, Adjusted maximum-likelihood method; Empirical best linear unbiased prediction; Fay--Herriot model; Linear mixed model; Mean squared error.

\section{Introduction}
\indent In recent decades, there has been high demand for reliable statistics on smaller geographic areas and sub-populations where large samples are not available. Considering the limited number of observations, a design-based direct estimator is not reliable for such ``small areas''---as they are called. An empirical best linear unbiased prediction (EBLUP) estimator is widely used as an efficient estimator based on a specific linear mixed model. It would be quite interesting to use the mean squared error (MSE) of EBLUP as a measure of its uncertainty. For small-area inference, its MSE needs to be estimated with high accuracy since it is not generally derived in a closed form.
Given a consistent estimator of an unknown model variance parameter, the MSE of EBLUP is always larger than that of the best linear unbiased prediction (BLUP) estimator which assumes a known model variance parameter, under certain conditions (Kackar and Harville, 1984). In most small-area applications, sufficient accuracy cannot be achieved by ignoring this difference, which is of the order of $O(m^{-1})$ for large $m$ (number of areas). Moreover, the naive MSE estimator, a consistent estimator substituted for the model variance parameter in the MSE of BLUP, lacks second-order unbiasedness for sufficient asymptotic accuracy in small-area estimation with large $m$. 

\indent Therefore, several second-order unbiased MSE estimators, with some bias correction, are suggested in place of the naive estimator (Prasad and Rao, 1990; Datta and Lahiri, 2000; Butar and Lahiri, 2003; Das et al., 2004; Datta et al, 2005; Hall and Maiti, 2006; Li and Lahiri, 2010; Yoshimori and Lahiri, 2014).
In particular, Yoshimori and Lahiri (2014) achieved strictly positive variance estimation while maintaining a functional form of the second-order unbiased MSE estimator proposed in Datta and Lahiri (2000), by using the adjusted maximum-likelihood method.

\indent Incidentally, a relevant question that arises is, Can the naive MSE estimator provide second-order unbiasedness through the adjusted maximum-likelihood method?
To answer this question, this paper proposes a new method for naive MSE estimation as (\ref{NRE})-(\ref{M.N}) in Section 4, which achieves the desired asymptotic property while maintaining strict positivity. To obtain the result, Section 3 provides a theorem to choose a suitable adjusted maximum-likelihood method for a specified functional form of the second-order unbiased and strictly positive MSE estimator, and vice versa (to choose a suitable functional form of the second-order unbiased and strictly positive MSE estimator for a specified adjusted maximum-likelihood method).
Section 5 presents a performance comparison among certain MSE estimators, including ours. The regularity conditions and all technical proofs are deferred to the appendix.

\section{The uncertainty of EBLUP under the Fay--Herriot model}
\indent The Fay--Herriot model (Fay and Herriot, 1979) is widely used for small-area inference. For $i=1,\ldots,m,$
\begin{align}
\label{FH}
{\rm Level \ 1:}& y_{i}|\theta_i \stackrel{ind}{\sim} N(\theta_i,D_i);\notag\\
{\rm Level\ 2:}& \theta_i \stackrel{ind}{\sim} N(x^{\prime}_i\beta,A).
\end{align}
The level-1 model takes into account the sampling distribution of the direct estimator $y_i$ for the $i$th small area. The true small-area mean for the $i$th area, denoted by $\theta_i$, is linked to providing area-specific auxiliary variables $x_i=(x_{i1},\cdots,x_{ip})^{\prime}$ in the level-2 model. In practice, the coefficient vector $\beta$ in $\mathbb{R}^p$ and the model variance parameter $A$ in this linking model are unknown. The assumption of a known $D_i$ often follows from the asymptotic variances of the transformed direct estimates (Efron and Morris, 1975) or from empirical variance modelling (Fay and Herriot, 1979).
This model can be rewritten as a specific linear mixed model:
$$y_i=\theta_i+e_i=x_i^{\prime}\beta+u_i+e_i, \ i=1,\ldots, m,$$
where $u_i$ and $e_i$ are mutually independent with the normality assumption $u_i \stackrel{iid}{\sim}N(0,A)$ and $e_i\stackrel{ind}{\sim}N(0,D_i)$.
It is well known that among all linear unbiased predictors $\hat{\theta}_i$ of $\theta_i$, BLUP yields the minimum MSE, which is defined as $E[(\hat{\theta}_i-\theta_i)^2]$, where the expectation is defined with respect to the joint distribution of $y$ and $\theta$ under the Fay--Herriot model (\ref{FH}). 
We give the form of BLUP as follows:
$$\hat{\theta}_i^{B}=(1-B_i)y_i+B_i x^{\prime}_i\tilde{\beta},$$
where $B_i=\frac{D_i}{A+D_i}$ is called the shrinkage factor toward $x_i^{\prime}\tilde{\beta}$ from the direct estimator $y_i$ with $\tilde{\beta}=\tilde{\beta}(A)=( X^{\prime}V^{-1}X)^{-1} X^{\prime}V^{-1} y$, $y=(y_1\ldots,y_m)^{\prime}$, $X=(x_1,\ldots,x_m)^{\prime}$ and $V=diag \{A+D_1,\cdots,A+D_m\}$.

\indent Since $A$ is unknown in practice, the following EBLUP of $\theta_i$ is widely used for small-area inference, with $A$ replaced with its consistent estimator, $\hat{A}$, in $\hat{\theta}_i^{B}$:
$$\hat{\theta}_i^{EB}=(1-\hat{B}_i)y_i+\hat{B}_i x^{\prime}_i\hat{\beta},$$
where $\hat{B}_i=\frac{D_i}{\hat{A}+D_i}$ and $\hat{\beta}=\tilde{\beta}(\hat{A})$. Hereafter, the consistent estimator $\hat{A}$ also denotes an even-translation-invariant estimator for all $\beta$ and $y$ that achieve an unbiasedness in the EBLUP, as in Kackar and Harville (1981).
To estimate the model variance parameter $A$, we can use the method of moments estimator (Fay and Herriot, 1979; Prasad and Rao, 1990) and the standard maximum-likelihood estimators, such as the profile maximum-likelihood (PML) and the residual maximum-likelihood (REML) estimators.
In particular, the REML estimator of $A$ is preferred in terms of its higher-order asymptotic accuracy for large $m$. Let $\hat{A}_{RE}$ denote the REML estimator of $A$, obtained as
$$\hat{A}_{RE}=\argmax_{0\leq A<\infty} L_{RE}(A|y),$$
where the residual likelihood function is $$L_{RE}(A|y)=|X^{\prime}V^{-1}X|^{-1/2}|V|^{-1/2}\exp\{ -y^{\prime}Py/2\}$$ and $P=V^{-1}-V^{-1}X( X^{\prime}V^{-1}X)^{-1}X^{\prime}V^{-1}.$

\indent However, the REML estimator of $A$ has serious problem such that it could be zero when $m$ (number of small areas) is not large enough, even though $A=0$ is not realistic in the context of small-area estimation.
In order to avoid zero estimates, Li and Lahiri (2010) and Yoshimori and Lahiri (2014) suggested using the specific adjusted maximum-likelihood estimators. Let $\hat{A}_{LL}$ and $\hat{A}_{YL}$ denote the respective estimators, given as
\begin{align*}
\hat{A}_{ad}=\argmax_{0\leq A<\infty}\tilde{L}(A)L_{RE}(A),
\end{align*}
where $\hat{A}_{ad}\in \{\hat{A}_{LL},\hat{A}_{YL} \}$, and $\tilde{L}(A)$ are adopted from their specific adjustment factors, $\tilde{L}(A)=A$ for $\hat{A}_{LL}$ and $\tilde{L}(A)=\arctan[tr(I-B_*)]^{1/m}$ for $\hat{A}_{YL}$ with $B_*=diag(B_1,\ldots,B_m)$.

\indent The MSE of BLUP under the Fay--Herriot model can be derived in a closed form as
\begin{align*}
MSE_i[\hat{\theta}_i^{B}]&\equiv E\left[ (\hat{\theta}_i^{B}-\theta_i)^2\right]=g_{1i}(A)+g_{2i}(A),
\end{align*}
where $g_{1i}(A)=\frac{AD_i}{A+D_i}$ and $g_{2i}(A)=\frac{D_i^2}{(A+D_i)^2}x_i^{\prime}(X^{\prime}V^{-1}X)^{-1}x_i$.
Unlike BLUP, EBLUP cannot generally provide a closed-form MSE, so we need to estimate the MSE of EBLUP from observed data in order to measure the uncertainty of EBLUP.
One simple MSE estimator, called the naive MSE estimator, can be constructed by plugging $\hat{A}_{RE}$ into $A$ in $MSE_i^{BLUP}$:
\begin{align}
\tilde{M}^{N}_i[\hat{\theta}_i(\hat{A}_{RE})]= g_{1i}(\hat{A}_{RE})+g_{2i}(\hat{A}_{RE}),\label{M.RE.N}
\end{align}
where $\tilde{M}_i^{N}$ denotes the naive MSE estimator of EBLUP under the REML method.

However, Kackar and Harville (1984) showed that the MSE of BLUP is smaller than that of EBLUP because the term depends on the variability of the estimator for $A$, which is of the order of $O(m^{-1})$ for large $m$, and it is not accurate enough to be ignored for small-area inference (Prasad and Rao, 1990). The result also implies that the bias of $\tilde{M}_i^{N}(\hat{A}_{RE})$ is of order of $O(m^{-1})$ under certain regularity conditions.
To gain more efficiency even for such situations, Prasad and Rao (1990) obtained an approximation of true MSE, $MSE_i$, up to the order $O(m^{-1})$, and the second-order unbiased MSE estimator, $\hat{M}_i$, of EBLUP with a method of moments estimator of $A$, so as to satisfy $E[\hat{M}_i-MSE_i]=o(m^{-1})$ for large $m$. Datta and Lahiri (2000) and Das et al. (2004) provided such approximation and an MSE estimator of EBLUP with REML based on the Taylor linearization method:
\begin{align}
MSE_i^{RE}\equiv &MSE_i[\hat{\theta}_i^{EB}(\hat{A}_{RE})]=g_{1i}(A)+g_{2i}(A)+g_{3i}(A)+o(m^{-1});\label{MSE.RE}\\
\hat{M}_i^{DL}=& g_{1i}(\hat{A}_{RE})+g_{2i}(\hat{A}_{RE})+2g_{3i}(\hat{A}_{RE}),\label{M.DL}
\end{align}
where $g_{3i}(A)={2D_i^2}/[(A+D_i)^3 tr(V^{-2})]$ and $\hat{M}_i^{DL}$ is second-order unbiased under certain regularity conditions such that $E[\hat{M}_i- MSE_i^{RE}]=o(m^{-1})$.

As mentioned above, MSE estimators generally require some bias correction methods to provide second-order unbiasedness.

\section{General functional form of MSE estimation for achieving second-order unbiasedness and strict positivity}

As in Hirose (2016), we consider the general functional form of an MSE estimator, denoted as: 
$$\hat{M}_i^g(\hat{A}_i)= g_{1i}(\hat{A}_{i})+g_{2i}(\hat{A}_{i})+c_i(\hat{A}_{i})g_{3i}(\hat{A}_{i}),$$ with some function $c_i(A)$, where $\hat{A}_i$ is a general adjusted maximum-likelihood estimator, defined as 
$$\hat{A}_i=\arg\max_{0\leq A<\infty}\tilde{L}_i(A)L_{RE}(A|y),$$ with the general adjustment factor $\tilde{L}_i(A)$, satisfying Condition A1 given in the appendix.

We also present a theorem on how to select an adjustment factor, $\tilde{L}_i(A)$, for the specified functional form of a second-order unbiased and strictly positive MSE estimator using the adjusted maximum-likelihood method. This theorem also comes in handy to choose a suitable functional form of the second-order unbiased and strictly positive MSE estimator for a specified adjusted maximum-likelihood method.

\begin{thm}
\label{th1}
{\rm
Under the regularity conditions and Condition A1, when we use $c_i(A)$ for the adjustment factor $\tilde{L}_i(A)$, such that
\begin{align}
\frac{\partial \log \tilde{L}_{i}(A)}{\partial A}=\frac{2-c_i(A)}{(A+D_i)}+o(1),\label{diff.eq}
\end{align}
with $c_i(A)\leq2$ being of the order of $O(1)$ for large $m$, satisfying $$\frac{\partial c_i(A)}{\partial A}(A+D_i)-c_i(A)+2\geq 0,$$
the following results hold:
\begin{description}
\item[(i)]
$MSE_i^g\equiv MSE[\hat{\theta}_i^{EB}(\hat{A}_i)]= g_{1i}({A})+g_{2i}({A})+g_{3i}({A})+o(m^{-1}),\\
\ \ \ =MSE_i^{RE}+o(m^{-1})$;
\item[(ii)]$E[\hat{M}_i^g(\hat{A}_i)]=MSE_i^g+o(m^{-1})$;
\item[(iii)]With additional condition $c_i(A)\geq 0$, we obtain $\hat{M}_i^g(\hat{A}_i)>0$;
\item[(iv)]There exists at least one estimate $\hat{A}_i^S$ for $A>0$, with the conditions $c_i(A)\geq0$ and $m>p+4$. 
A more progressive existence condition is required for the number of areas $m>p$ such that $\exp\left[\log A^2-\int \frac{c_i(A)}{A}d A\right]=o(A^{(m-p)/2})$ holds for large $A$,
\end{description}
where $\hat{M}_i^g(\hat{A}_i)= g_{1i}(\hat{A}_{i})+g_{2i}(\hat{A}_{i})+c_i(\hat{A}_{i})g_{3i}(\hat{A}_{i})$ and $$\hat{A}_i^S=\arg\max_{0\leq A<\infty}\tilde{L}_i(A)\tilde{L}_{add}(A)L_{RE}(A|y)$$ with $\tilde{L}_{add}(A)$ satisfying Condition A2-A3. Incidentally, even if $\hat{A}_i$ replaces to $\hat{A}_i^{S}$, parts (i)-(iii) hold. 
If $\tilde{L}_{i}(A)\Big|_{A=0}=0$ holds, we no longer need to consider the $\tilde{L}_{add}(A)$ term. 
 
}
\end{thm}
The proof of Part (i) can be obtained along the same lines as in Das et al. (2004). The proof of Part (iii) follows from the definition of $\hat{M}^{g}_i$. Parts (ii) and (iv) are deferred to the appendix. 
Thus, $c_i(A)$ should move only between 0 and 2.

\begin{col}
{\rm
From Theorem \ref{th1}, when the REML estimator is used, the value $2$ is selected as a suitable $c_i(A)$, corresponding to $\hat{M}_i^{DL}$.

}
\end{col}

\section{Second-order unbiased naive MSE estimator}

Theorem \ref{th1} ensures that the naive MSE estimator provides second-order unbiasedness and strict positivity by setting $c_i(A)=0$. Thus, we obtain a suitable adjustment factor, $\tilde{L}_i^{N}(A)$, up to the order of $O(1)$ for large $m$, after solving $\frac{\partial \log \tilde{L}_i(A)}{\partial A}=\frac{2}{(A+D_i)}+o(1)$:
\begin{align*}
\tilde{L}_i^N(A)=C(A+D_i)^2,
\end{align*}
where $C$ is a generic positive constant.

However, the estimates could be zero since $\tilde{L}_i^N(A)\mid_{A=0}\neq 0$, as described in Yoshimori and Lahiri (2014).
To avoid this problem, we add an additional adjustment factor, $\tilde{L}_{add}(A)$, satisfying Conditions A2-3. For example, $\tilde{L}_{add}(A)$ can be adopted as the specific adjustment factor as in Yoshimori and Lahiri (2014).
Thus, we finally obtain the specific adjusted maximum-likelihood estimator, denoted by $\hat{A}_i^{N}$, to construct the second-order unbiased naive MSE estimator while maintaining strict positivity from Theorem \ref{th1}:
\begin{align}
\hat{A}_i^{N}=\arg\max_{0\leq A<\infty}\tilde{L}_i^{N}(A)\tilde{L}_{add}(A)L_{RE}(A|y). \label{NRE}
\end{align}

Let $\hat{M}^{N}$ denote the new naive MSE estimator:
\begin{align}
\hat{M}_{i}^{N}=g_{1i}(\hat{A}_{i}^N)+g_{2i}(\hat{A}_{i}^N).\label{M.N}
\end{align}
Additionally, we also show the result such that ${MSE}_{i}[\hat{\theta}_i(\hat{A}_{i}^N)]=MSE_i^{RE}+o(m^{-1})$ from Theorem \ref{th1} (i). 

Next, we obtain the following theorem on the properties of $\hat{A}_{i}^N$.
\begin{thm}
\label{bias.A}
{\rm Under the regularity conditions and Conditions A2-3, we have, for large $m$,
\begin{description}
\item[(i)] $E[\hat{A}_{i}^N-A]=\frac{4}{tr[V^{-2}](A+D_i)}+o(m^{-1})$,
\item[(ii)] $E[(\hat{A}_{i}^N-A)^2]=\frac{2}{tr[V^{-2}]}+o(m^{-1})$,
\item[(iii)] $\hat{A}_{i}^N$ is strictly positive for $m>p+4$. 
\end{description}
}
\end{thm}
The proofs of parts (i) and (ii) are similar to those shown in Yoshimori and Lahiri (2014).
For Part (iii), the proof follows from Theorem \ref{th1} (iv) by setting $c_i(A)=0$.

\section{Simulation study}
\indent In this section, we compare performances among different estimators of both the variance parameter $A$ and the MSE of the EBLUP, mentioned in the previous section.
In order to investigate the effect of $m$ and $B_i$, we assume that $m=15$ in a balanced case such that $B_i=B$ patterns: \{0.1,0.3,0.5,0.7,0.9\} with fixed $D_i=D=1$ for all areas.
We generated $10^{4}$ independent data sets $\{y_i,\ i=1,\ldots, m\}$ from the Fay--Herriot model (\ref{FH}) with $x_{i}^{\prime}\beta=0$ and $p=1$. In this simulation study, we also estimate this zero mean from a practical perspective. 
In terms of MSE evaluation, we compared the MSE of the EBLUP with two different estimators: REML $\hat{A}_{RE}$ and our new estimator $\hat{A}^N$. We denote them as \lq\lq{}REML\rq\rq{} and \lq\lq{}NRE\rq\rq{}, respectively. When the REML yielded zero estimates, we treated them as 0.01. 

Table \ref{MSE} shows each simulated MSE of the EBLUP multiplied by $100$, based on two variance estimation methods. From this result, the new variance estimator provides very similar performance to REML in terms of MSE of EBLUP for small or moderate $B$ values. In contrast, the new variance estimator does not achieve better performance than the REML method for large $B$ values in terms of MSE.

\begin{table}[h]
\centering
\caption{MSE of EBLUP based on REML and NRE methods, multiplied by 100 } 
\label{MSE}
\begin{tabular}{r|rrrrr}
  \hline
$B_i=B$ & 0.1 & 0.3 & 0.5 & 0.7 & 0.9 \\ 
  \hline
RE & 92.26 & 77.17 & 59.83 & 40.27 & 20.66 \\ 
  NRE & 92.26 & 77.19 & 60.97 & 44.69 & 28.35 \\
   \hline
\end{tabular}
\end{table}
Note: The table shows values increased 100-fold to allow easy comparison.

We also report the percentage of the relative biases (PRB) of different MSE estimators for the MSE of EBLUP with REML in Table \ref{RB.tb} considering the good performance of the MSE of EBLUP with REML, shown in Table 1. 
PRB is defined as
$$PRB: \ \frac{\hat{M}_i-MSE_i^{RE}}{MSE_i^{RE}}\times 100,$$
where $\hat{M}_i$ denotes an MSE estimator and $MSE_i^{RE}$ is defined as in (\ref{MSE.RE}). 
We now consider three MSE estimators for $\hat{M}_i$, $\tilde{M}_i^{N}(\hat{A}_{RE})$, $\hat{M}_i^{DL}(\hat{A}_{RE})$, and $\hat{M}_i^{N}(\hat{A}^{N})$, defined in (\ref{M.RE.N}), (\ref{M.DL}), and (\ref{M.N}). Hereafter, we denote them as \lq\lq{}Naive.RE\rq\rq{}, \lq\lq{}DL.RE\rq\rq{}, and \lq\lq{}Naive.N\rq\rq{}.

\begin{table}[h]
\centering
\caption{Percent RB (PRB) of MSE estimates for MSE of EBLUP with REML} 
\label{RB.tb}
\begin{tabular}{r|rrr}
  \hline
$B_i=B$ & Naive.RE & DL.RE & Naive.N \\ 
  \hline
0.1 & -3.48 & -0.08 & -0.08 \\ 
0.3 & -12.65 & -0.57 & -0.63 \\ 
0.5 & -21.29 & 3.99 & 2.78 \\ 
0.7 & -22.56 & 26.27 & 19.23 \\ 
0.9 & -3.29 & 107.40 & 75.57 \\ 
   \hline
\end{tabular}
\end{table}

From the table, the naive estimator with REML (Naive.RE) tends to be underestimated, unlike other estimators. 
It probably occurs from the absence of a positive bias correction term to achieve second-order unbiasedness. 
As regards other estimation methods, the performance of the second-order unbiased naive MSE estimator is similar to that of DL.RE for small and moderate $B$ values. 
Moreover, our naive estimator, $\hat{M}_i^{N}(\hat{A}^N)$, remedies the over-estimation issue caused by DL.RE for large $B$ values. 

\section{Conclusion}
\indent In this paper, we established that the new estimator $\hat{A}_{i}^N$ conduces to a second-order unbiased naive MSE estimator while maintaining strict positivity. Results show that the new method remedies the under-estimation issue associated with the existing naive estimator.
Moreover, we also revealed the relationship between the general functional form of MSE estimation and the general adjustment factor $\tilde{L}_{i}(A)$. 
Consequently, we can, on the one hand, easily construct a second-order unbiased and strictly positive MSE estimator for EBLUP using the specified adjusted maximum-likelihood method and, on the other, select an adjustment factor with the above MSE estimator. 

\section*{Acknowledgement}
The author\rq{}s research was supported by Grant-in-Aid for Research Activity start-up, JSPS Grant Number 26880011.

\appendix
\def\thesection{Appendix.\Alph{section}}

\section{Conditions}
\subsection*{Regularity Conditions}

We assume the following regularity conditions:
\begin{description}
\item [R1] $rank(X)=p$ is fixed for large $m$;
\item [R2]The elements of $X$ are uniformly bounded such that $\sup_{i\geq 1}h_{ii}=O(m^{-1})$, where $h_{ii}=x_{i}^{\prime}(X^{\prime}X)^{-1}x_{i}$;
\item [R3]$0<\inf_{i\geq 1}D_i\leq \sup_{i\geq 1}D_i<\infty$, \ $0<A<\infty$;
\item [R4]$|\hat{A}_{i}|<C_{ad}m^{\lambda}$, where $C_{ad}$ is a generic positive constant and $\lambda$ is a small positive constant, where $\hat{A}_{i}$ is a general adjustment maximum-likelihood estimator of $A$.
\end{description}

We also consider the class of adjustment factors $\tilde{L}_i(A)$ and $\tilde{L}_{add}(A)$ as
\subsection*{Conditions A}
\begin{description}
\item [A1] $\log \tilde{L}_i(A)$ is independent of $y$ and four times continuously differentiable with respect to $A$ and a strictly monotonically increasing and concave function of $A>0$. Moreover, $\frac{\log \tilde{L}_{i}(A)} {\partial A^k}$ is of the order of $O(1)$ for large $m$ with $k=0,1,2,3,4$;
\item [A2]$\log \tilde{L}_{add}(A)$ is independent of $y$ and four times continuously differentiable with respect to $A$. Moreover, $\frac{\partial^k \log \tilde{L}_{add}(A)} {\partial A^k}$ is of the order of $o(1)$ for large $m$ with $k=0,1,2,3,4$;
\item [A3]$\log \tilde{L}_{add}(A)$ is a strictly monotonically increasing and concave function of $A>0$ with $\tilde{L}_{add}(A)\Big|_{A=0}=0$ and  $\tilde{L}_{add}(A)<C$ on $A>0$ with a generic positive constant $C$.
\end{description}

\section{Proof of Theorem \ref{th1} (ii) and (iv)}
\subsection{Proof of Theorem \ref{th1} (ii)}
From Theorem1 on Yoshimori and Lahiri (2014), we have for large $m$
\begin{align}
E[g_{1i}(\hat{A}_i)-g_{1i}(A)]=&B_i^2\frac{2}{tr[V^{-2}]}\frac{\partial \log \tilde{L}_{i}(A)}{\partial A}-g_{3i}(A)+o(m^{-1}).\label{g.bias}
\intertext{Using Theorem \ref{th1} (i) and the result (\ref{g.bias}),}
E[\hat{M}_i^g-MSE_i^{g}]=&E[g_{1i}(\hat{A}_i)+g_{2i}(\hat{A}_i)+c_i(\hat{A})g_{3i}(\hat{A}_i)]\notag\\
&-[g_{1i}({A})+g_{2i}({A})+g_{3i}({A})]+o(m^{-1});\notag\\
=&B_i^2\frac{2}{tr[V^{-2}]}\frac{\partial \log \tilde{L}_{i}(A)}{\partial A}+(c_i(A)-2)g_{3i}(A)+o(m^{-1}). \label{g.Mhatbias}
\end{align}

If the first two terms on the right-hand side of (\ref{g.Mhatbias}) vanish for second-order unbiasedness, we obtain the following differential equation:
\begin{align*}
\frac{\partial \log \tilde{L}_{i}(A)}{\partial A}&=\frac{[2-c_i(A)]}{A+D_i}+o(1).
\end{align*}
Thus, Theorem \ref{th1} (ii) follows.

\subsection{Proof of Theorem \ref{th1} (iv)}
We shall first prove with regard to the progressive condition for $\hat{A}_i^G$ existence on $A>0$. 

From Conditions A1-A3, we have for $A> 0$
$$\tilde{L}_{i}(A) \tilde{L}_{add}(A)L_{RE}(A)\Big{|}_{A=0}=0, \ \ {\rm and} \ \ \tilde{L}_{i}(A) \tilde{L}_{add}(A)L_{RE}(A)\Big{|}_{A=0}>0.$$
Thus, it suffices to show that for large $A$,
\begin{align}
\tilde{L}_{i}(A) \tilde{L}_{add}(A)L_{RE}(A)=o(1).\label{exist}
\end{align}

Let $C$ denote a generic positive constant.
From the fact that
$$L_{RE}(A)<C(A+\sup_{i\geq 1} D_{i})^{\frac{p}{2}}| X^{\prime}X |^{-\frac{1}{2}}(A+\inf_{i\geq 1}D_{i})^{-\frac{m}{2}},$$
Condition A3 reduces (\ref{exist}) to the following, for large $A$:

\begin{align*}
\tilde{L}_{i}(A)=o(A^{(m-p)/2}).
\end{align*}
The solution of the differential equation (\ref{diff.eq}), $\tilde{L}_i(A)=C\exp\left[\log(A+D_i)^2-\int\frac{c_i(A)}{(A+D_i)}d A \right]$, further reduces to
\begin{align}
\exp\left[\log A^2-\int\frac{c_i(A)}{A}d A \right]=o(A^{(m-p)/2}).\label{exist2}
\end{align}
Thus, we need the number of areas $m>p$ to satisfy the above condition (\ref{exist2}) with fixed $p$ and $c_i(A)$ for Conditions A1-A3. 
When $c_i(A)\geq0$ holds, (\ref{exist2}) reduces to 
$$\exp\left[\log A^2-\int\frac{c_i(A)}{A}d A \right]\leq A^2. $$
Hence, $m>p+4$ can be a conservative existence condition for $\hat{A}_i^{G}$ for $A>0$.

\end{document}